\RequirePackage{rotating}
\documentclass[a4paper,10pt]{iopart}
\usepackage{iopams}  
\usepackage[colorlinks,citecolor=blue,linkcolor=red,anchorcolor=blue,filecolor=blue,urlcolor=blue]{hyperref}
\usepackage{color}
\usepackage{graphicx}
\usepackage{cite}
\usepackage{rotating}
\usepackage{afterpage}

\begin{document}
\title{Indispensability of cross-shell contributions in neutron resonance spacing}
\bigskip
\author{T. Ghosh$^{1,2,\star}$, Sangeeta$^3$, B. Maheshwari$^{4}$, G. Saxena$^5$, B. K. Agrawal$^{1,2}$}

\address{$^1$Saha Institute of Nuclear Physics, Kolkata-700064, India}
\address{$^2$Homi Bhabha National Institute, Anushakti Nagar, Mumbai-400094, India}
\address{$^3$Department of Physics, University Institute of Sciences, Chandigarh University, Gharuan, Mohali-140413, Punjab, India}
\address{$^4$Department of Physics, Faculty of Science, University of Zagreb, HR-10000 Zagreb, Croatia}
\address{$^5$Department of Physics (H $\&$ S), Govt. Women Engineering College, Ajmer-305002, India}

\ead{ghoshtanmoy536@gmail.com$^\star$}

\begin{abstract}
\noindent
Spin and parity dependent nuclear level densities (NLDs) are obtained for configuration interaction shell model using a numerically efficient spectral distribution method. The calculations are performed for $^{24}$Na, $^{25,26,27}$Mg nuclei using full $sd$-$pf$ model space that incorporates the cross-shell excitations from $sd$ to $pf$-shell. The NLDs so obtained are then employed to determine the s-wave neutron resonance spacing (D$_0$) which is one of the crucial inputs for the predictions of astrophysical reaction rates. Though the considered nuclei are not neutron-rich, the contributions from cross-shell excitations to $pf$-shell are indispensable to explain the experimental data for D$_0$ which otherwise are significantly overestimated.
\end{abstract}

\section{Introduction} 
Nuclear level densities (NLDs) are one of the key ingredients for studying the nuclear reactions \cite{Hauser1952}, particularly the astrophysical reaction rates at a fixed temperature in terms of Maxwellian average of cross-sections over a wide range of energy \cite{claus1989,horoi2011,Nobre2020, Martinez2022,Sangeeta2022}. The NLDs have been calculated using many different approaches such as simple phenomenological models based on non-interacting degenerate Fermi gas \cite{Gilbert1965,Dilg1973,Ignatyuk1993,Rajasekaran1982} and semi-classical models \cite{Kaur2015,Dwivedi2019} to more complex microscopic mean-field models \cite{Goriely2008,Ghosh2022,Hung2020}. The collective effects are included through the rotational and vibrational enhancement factors in these models. The NLDs in mean-field models at low excitation energies are computed using combinatorial method \cite{Egorov1989,Egorov1989npa} which are further normalized with the experimental data at low energy and neutron resonance energy. The statistical method is used at higher excitation energies. \par

More realistic values of NLDs are obtained using the framework of shell model which incorporates the effects of collective excitations naturally through the configuration mixing due to the residual interaction \cite{Coraggio2009}. The shell model Monte Carlo \cite{Johnson1992,Lang1993,nakada1997,nakada1998,Nakada2008,Alhassid2015,Alhassid2015EPJA,Alhassid2016,Mustonen2018} utilizes auxiliary fields to compute the thermal trace for the energy and further inverse Laplace transform to obtain the NLDs. An accurate estimation of the NLDs described within the configuration space are also obtained using the extrapolated Lanczos method \cite{Ormand2020}. 
An efficient way to construct the NLDs is based on the spectral distribution method (SDM) \cite{Chang1971,French1971,Kota1989,Kota1996,Horoi2003,Horoi2004,Jacquemin1981} for many-body shell model Hamiltonian in full configuration space, which avoids the diagonalization of huge dimensional matrices. This method allows one to incorporate the many-body effects on the wave functions appropriately and is a basis for the applications of statistical spectroscopy generated by many-body chaos \cite{kar1994,Kota1995,Gomez2011,Kota2018}; see Ref. \cite{Zelevinsky2019} for a recent review. 
The SDM facilitates the calculations of first and second moments of the Hamiltonian for different configurations at fixed spin and parity, required to construct the NLDs \cite{Senkov2010,Senkov2013,Senkov2016}. It has also been extended to calculate non-spurious spin- and parity-dependent NLDs for the valance spaces with more than one major harmonic oscillator shells \cite{Senkov2011plb}. These NLDs agree reasonably with the exact shell model calculations \cite{Senkov2013}.  \par

The values of s-wave neutron resonance spacing ($D_0$) which depend inversely on the NLDs for a specific spin at neutron separation energy play crucial role in estimating neutron capture reaction rates of astrophysical interest \cite{Ormand2020,Sangeeta2022}. The success of the phenomenological and microscopic mean-field models in explaining the experimental astrophysical reaction rates relies to some extent on scaling the NLDs to the measured $D_0$ values. The SDM has been applied to construct full database of total and spin dependent NLDs for all $sd$-shell nuclei \cite{Karampagia2018}. The calculations are performed with $sd$ model space which can produce NLDs for only positive parity levels. The NLDs for $sd$ model space yield significantly smaller $D_0$ which suggests that the NLDs are underestimated. The contributions from cross-shell excitations by including the orbitals either from core or from higher-lying shells may enhance the NLDs and lead to appropriate values of $D_0$. \par
The importance of cross-shell excitations between $sd$-shell to $pf$-shell has been successfully explored by performing systematic study of two neutron separation energies, excitation energies and B(E2) values in Ne, Mg and Si isotopes \cite{utsuno1999}. The shape transitions in the neutron-rich exotic Si and S isotopes are explained on the basis of tensor-force nature of the Hamiltonian by developing SDPF-MU interaction \cite{Utsuno2012}. The shell model calculations for neutron-rich S isotopes could explain the available experimental level scheme, electric quadrupole and the magnetic dipole moments, and spectroscopic factors using $sd$-$pf$ model space \cite{saxena2017}. A detailed theoretical shell model investigation in $sd$ and $sd$-$pf$ model spaces for $^{28}$Mg highlighted the significance of $pf$-shell contributions for explaining three higher energy levels beyond 7 MeV \cite{williams2019}. Recently, the location of neutron-drip line in F and Ne isotopes has been analyzed where the small occupancies in $pf$-shell are found to alter the results significantly, underlining the need of cross-shell excitations \cite{liu2021}. 

In this article, we present an improved calculation of the realistic NLDs for few $sd$-shell nuclei using SDM
in $sd$-$pf$ model space to obtain s-wave neutron resonance spacing. This significantly extends previous shell-model studies performed in the $sd$-shell \cite{Karampagia2018}, increasing the total number of single-particle orbitals from six to fourteen. The calculations have been performed for $^{24}$Na, $^{25,26,27}$Mg nuclei near the line of stability for which the experimental $D_0$ values are available. For these nuclei, the $sd$-$pf$ model space as considered may be sufficient to explore the excitations near neutron separation energies. To the best of our knowledge, the NLDs and corresponding neutron resonance spacing for $^{24}$Na, $^{25,26,27}$Mg nuclei near the line of stability have not been investigated using SDM in $sd$-$pf$ model space so far.



\section{Spectral Distribution Method}
The NLDs are obtained using spectral distribution method \cite{Senkov2013} applied to shell model Hamiltonian with a realistic residual interaction. One calculates first and second moments of Hamiltonian for the full configuration space within the spectral distribution method. The NLDs are thus obtained using the Gaussian distribution of the levels constructed by these moments. For a given nucleus, the valence nucleons are distributed in many ways over available orbitals within the configuration space. Each resulting configuration, or the partition, contains $D_{\alpha p}$ many-body states with $\alpha$ being exact quantum numbers and $p$ is the partition. The states present in a given partition are distributed over some energy region as a result of interactions inside the partition. For each partition, the statistical average of an operator $\hat{O}$ over the states is defined as,
\begin{eqnarray}
\langle \hat{O} \rangle_{\alpha p} = \frac{1}{D_{\alpha p}} Tr^{\alpha p} \hat{O} 
\end{eqnarray}
In particular, the centroid energy of the partition is the first moment of the Hamiltonian,
\begin{eqnarray}
E_{\alpha p} = \frac{1}{D_{\alpha p}} Tr^{\alpha p} \hat{H}
\end{eqnarray}
This comes directly from the diagonal elements of the Hamiltonian matrix. The second moment of the Hamiltonian,  
\begin{eqnarray}
\sigma^2_{\alpha p} = \langle H^2 \rangle_{\alpha p} -E^2_{\alpha p} = \frac{1}{D_{\alpha p}} Tr^{\alpha p} H^2 -E_{\alpha p}^2
\end{eqnarray}
is determined by the off-diagonal elements of the Hamiltonian matrix including the interaction between partitions. Explicit diagonalization of the full Hamiltonian is not required now, since the first and second moments can be directly obtained from the matrix representation. The actual distributions can be very well approximated to the Gaussians which is a manifestation of quantum complexity and chaotization \cite{Mon1975,Brody1981,Wong1986,Kota2010}. Hence, the level density $\rho(E; \alpha)$ is obtained as
\begin{eqnarray}\label{rho}
\rho(E; \alpha)=\sum_p D_{\alpha p} G_{\alpha p}(E)
\end{eqnarray}
where $G_{\alpha p}(E)$ is the Gaussians weighted over their dimensions for all partitions at given energy $E$. Finite range or truncated Gaussians can provide appropriate results for each partition, 
\begin{eqnarray}
G_{\alpha p}(E)= G(E - E_{\alpha p} + E_{\rm gs}; \sigma_{\alpha p}) \label{egs}.
\end{eqnarray}
The Gaussians are cut off at a distance $\approx 2.3 \sigma_{\alpha p}$ from the corresponding centroid and then re-normalized \cite{Senkov2011plb,Senkov2013} to remove unphysical tails. The ground state energies $E_{\rm gs}$ are to be calculated using full Hamiltonian matrix for being consistent with the first and second moments. In the M-scheme, the level density $\rho(E;J)$ for certain spin $J$ can be calculated by the difference of $\rho(E;M=J)$ and $\rho(E;M=J+1)$.\par

We calculate the ground state energies $E_{\rm gs}$ appearing in Eq. (\ref{egs}) for $^{24}$Na, $^{25,26,27}$Mg nuclei in $sd$-$pf$ model space. The shell model Hamiltonian has been diagonalized for SDPF-MU effective interaction \cite{Utsuno2012} using KSHELL \cite{Shimizu2019}. The SDPF-MU interaction is known to describe well the shell evolution and the spectroscopy of neutron-rich nuclei in the upper $sd$-shell. The $sd$-part of this interaction is based on the well known USD interaction \cite{Brown1988} whereas the $pf$-part is based on the GXPF1B interaction \cite{Honma2009}, which explains very successfully the spectroscopy of $pf$-shell nuclei. The GXPF1B Hamiltonian was created from the GXPF1A Hamiltonian \cite{Honma2005} by changing five $T=1$ matrix elements and the single-particle energies (SPE) which involved 2$p_{1/2}$ orbital. By following USD and GXPF1B, all two-body matrix elements are scaled by a mass dependent factor $A^{-0.3}$. The SPE for the $sd$-shell are taken from USD, and those for the $pf$-shell are the effective SPE on top of the $^{40}$Ca closed-shell equal to the SPE of GXPF1B. The active proton and neutron orbitals are $1d_{5/2}$, $1d_{3/2}$, $2s_{1/2}$, $1f_{7/2}$, $1f_{5/2}$, $2p_{3/2}$, $2p_{1/2}$ with the energies of $-3.95, 1.65, -3.16, 5.06, 10.02, 1.18, 1.90$ MeV, respectively. After obtaining $E_{\rm gs}$, we have calculated the level density from Eq. (\ref{rho}) by employing Moments-Method (MM) code \cite{Senkov2011plb} which removes spurious excitation due to center of mass when the valence space contains more than one major harmonic oscillator shells.

\begin{table}[!htb]
\caption{\label{tab:egs} The shell model ground state energies $E_{gs}$ along with the ground-state spins for SDPF-MU interaction. The calculations are performed using (0,2)$\hbar\omega$ and (0,4)$\hbar\omega$ excitations to $pf$-shell except the case of $^{24}$Na and $^{25}$Mg where full space calculations are possible.}
\centering
\resizebox{0.6\textwidth}{!}{%
\centering
\begin{tabular}{c c c c c c}
\hline
Nucleus & J$^\pi$ & \multicolumn{3}{c}{$E_{gs}$ (MeV)}\\
\cline{3-5}
 &  & (0,2)  & (0,4)& full \\
\hline
$^{24}$Na & $4^+$ & $-83.31$ & $-83.99$ &$-84.00$ \\
$^{25}$Mg & ${5/2}^+$ &  $-102.03$ & $-103.44$ &$-103.56$ \\
$^{26}$Mg & $0^+ $  &  $-113.75$  & $-115.39$ & - \\
$^{27}$Mg & ${1/2}^+$ &  $-119.52$ & $-120.84$  & -\\
\hline
\end{tabular}}
\end{table}

\section{Result and discussion}
We calculate total NLDs for $^{24}$Na and $^{25,26,27}$Mg nuclei within the SDM approach with the inputs in terms of E$_{gs}$ and SDPF-MU interaction. Though the considered nuclei are not neutron-rich, it would be interesting to investigate the effect of the cross-shell excitations on the neutron resonance spacing. 
The required values of E$_{gs}$ along with ground state spins, as listed in Table \ref{tab:egs}, for these nuclei are obtained with (0,2)$\hbar\omega$ and (0,4)$\hbar\omega$ excitations to $pf$-shell because of the limited computational power. Here, (0,$n$)$\hbar\omega$ refer the allowed $pf$-shell excitations upto $n$ particles-$n$ holes.
The E$_{gs}$ for $^{24}$Na and $^{25}$Mg are also obtained with full $sd$-$pf$ model space calculations. The results for (0,4)$\hbar\omega$ excitations and full space calculations are very close to each other. We evaluate NLDs corresponding to $E_{gs}$ for (0,4)$\hbar\omega$ excitations in $pf$-shell.\par

\begin{figure*}[!htb]
\begin{center}
\includegraphics[width=5.0in]{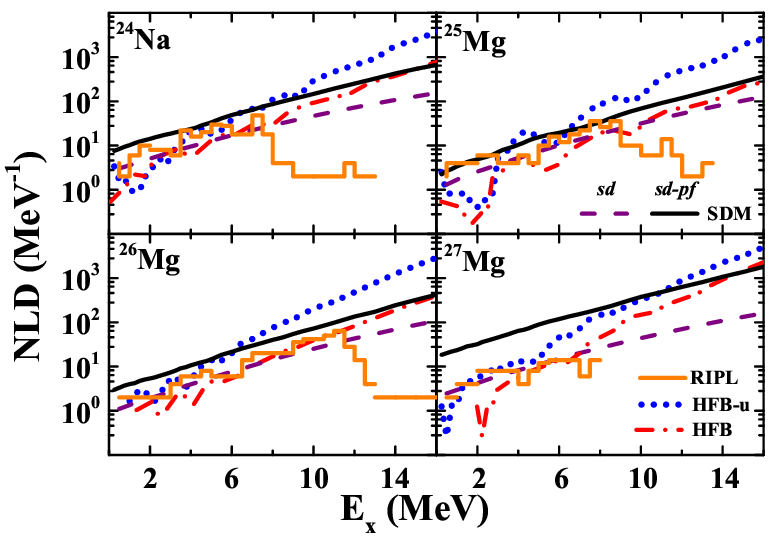}

\caption{Excitation energy ($E_x$) dependence of NLDs for $^{24}$Na and $^{25,26,27}$Mg obtained by employing SDM within $sd$-$pf$ model space. For comparison, we also show SDM results using $sd$ model space \cite{Karampagia2018} along with those for the mean-field models, HFB-u (un-normalized) and HFB (normalized) \cite{Goriely2008}. Histograms represent the NLDs obtained from low-lying discrete levels \cite{Capote2009}.} \label{nld}
\end{center}
\end{figure*}
Fig. \ref{nld} presents excitation energy ($E_x$) variation of nuclear level densities. These $sd$-$pf$ NLDs come closer to the experimental data of the low-lying discrete levels taken from RIPL \cite{Capote2009} in comparison to the NLDs calculated within $sd$ model space \cite{Karampagia2018} except for $^{27}$Mg. The NLDs obtained using the mean-field models, Hartree-Fock-Bogoliubov \cite{Goriely2008} HFB-u (unnormalized) and HFB (normalized), are also shown for comparison. The HFB results are in overall agreement with the experimental data due to the normalization with
low-lying energy levels and neutron resonance data. In the $sd$-model space, the considered nuclei exhibit dimensions in the order of a few thousand for the given spin and parity level. However, when we consider a larger
$sd$-$pf$ model space, these dimensions experience a significant increase, reaching the order of a few million, owing to a substantial rise in the number of available configurations for the given spin and parity level. Consequently, the nuclear level density, representing the number of levels per energy interval, increases and approaches closer to experimental data. The agreement observed in the nuclear level densities from the shell model for the $sd$-$pf$ model space underscores the crucial role played by the larger model space and the corresponding increase in available configurations and configuration mixing by cross-shell excitations.
The configuration mixing in the shell model naturally accounts for the collective excitations important in fine nuclear structure and are controlled by the amount of residual interactions which is absent in the mean-field models due to the ignorance to many-body correlations. The unnormalized mean-field results, therefore, stay far from the experimental data as shown by HFB-u in Fig. \ref{nld}, a gap typically addressed by fitting to the experimental low-lying levels, shown by HFB results. The deviations of calculated NLDs with respect to the low-lying levels taken from RIPL beyond 7 MeV can be attributed to the lack of availability of experimental high-lying states.

\begin{figure*}[!htb]
\vspace{-0.3cm}
\begin{center}
\includegraphics[width=0.9\textwidth]{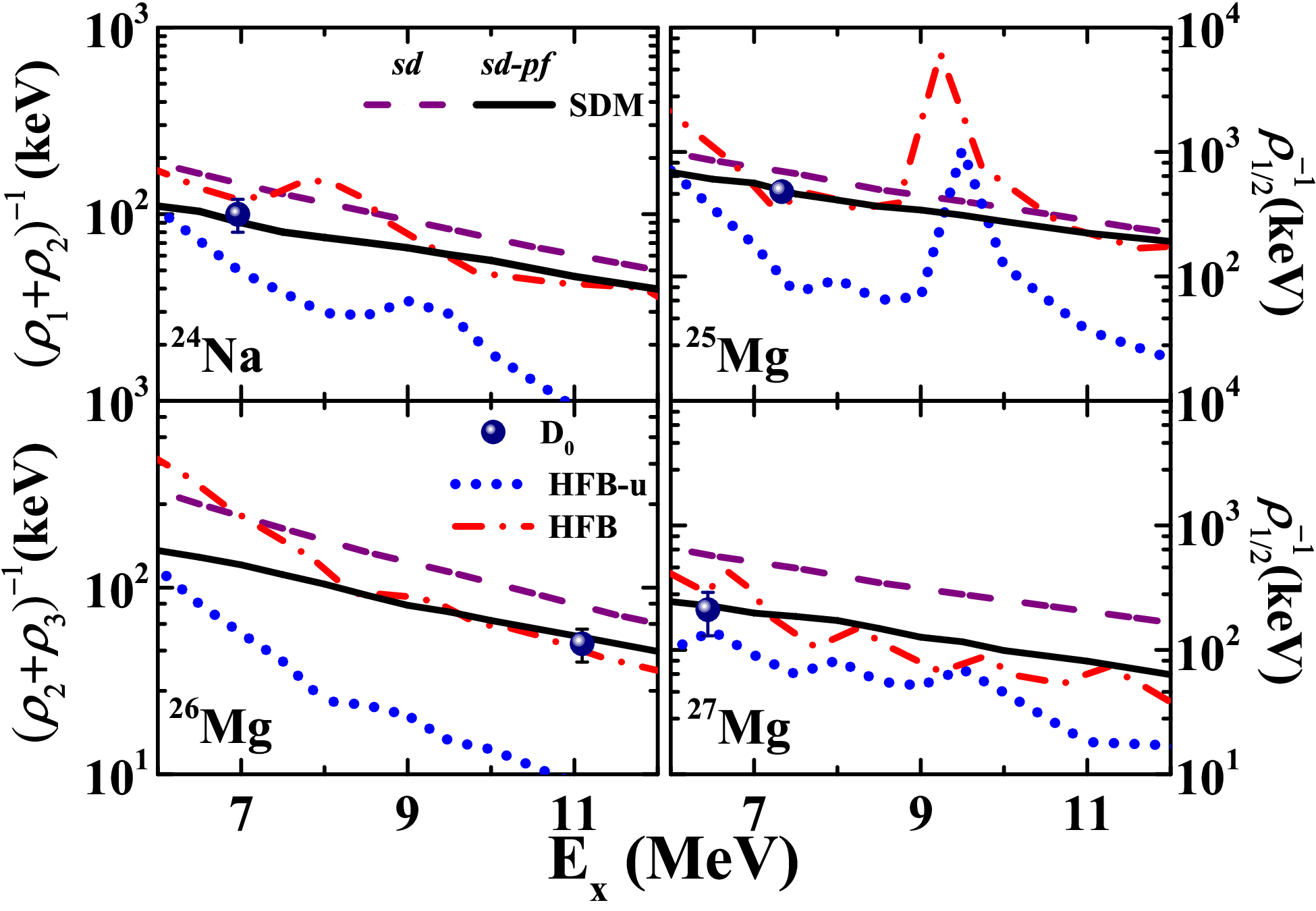}
\end{center}
\vspace{-0.3cm}
\caption{Inverse level density obtained from SDM in $sd$-$pf$ model space as a function of excitation energy for specific spins relevant for the estimation of s-wave neutron resonance spacing, ${D_0}$. The results are compared with the NLDs of $sd$ model space along with those of mean-field models, HFB-u (unnormalized) and HFB (normalized). Experimental datum for ${D_0}$ is also shown in each panel which matches well with the $sd$-$pf$ results.}\label{d0}
\end{figure*}

We calculate the neutron resonance spacing \cite{Gilbert1965} as 
\begin{eqnarray}
D_0 =\left\{\begin{array}{cc}
    \frac{1}{\rho(S_n, J_t + 1/2, \pi_t)+ \rho(S_n, J_t - 1/2, \pi_t)} \quad J_t \neq 0,   \\ 
    \\
\frac{1}{\rho(S_n, 1/2, \pi_t)} \quad J_t = 0; 
    \end{array} \right.
\label{D0}
\end{eqnarray}
where $S_{n}$ is the neutron separation energy, and $J_{t}$ and $\pi_t$ are the spin and parity of target nucleus, respectively. Fig. \ref{d0} presents the variation of inverse level density as a function of excitation energy in the range of $6-12$ MeV for specific spins relevant for the estimation of ${D_0}$. The $D_0$ values with $sd$-$pf$ model space are in remarkable agreement with the experimental data \cite{Capote2009}. These spacings play a crucial role in estimating neutron capture reaction rates of astrophysical interest \cite{Sangeeta2022}. Our results emphasize the role of cross-shell excitations from $sd$-shell to the $pf$-shell for the nuclei considered, though, they lie near the line of stability. For the comparison, we also show the results for $sd$ model space obtained using SDM and those for the mean-field models.\par

\begin{table}[!htb]
\caption{\label{tab:resonance} The s-wave neutron resonances spacing $(D_{0})$ obtained from the NLDs for $sd$ \cite{Karampagia2018} and $sd$-$pf$ model spaces using SDM. The experimental data \cite{Capote2009} and those obtained for mean-field models \cite{Goriely2008} are also presented for the comparison.}
\centering
\resizebox{0.7\textwidth}{!}{%
\begin{tabular}{c | c  c | r@{\hskip 0.1in} c @{\hskip 0.1in}c @{\hskip 0.1in}c@{\hskip 0.1in}  c}
\hline
\multicolumn{1}{c|}{Nucleus}&
\multicolumn{1}{c}{$J_t^{{\pi}_{t}}$}&
\multicolumn{1}{c|}{$S_{n}$}&
\multicolumn{5}{c}{$D_0$ (keV)}\\
\cline{4-8}
\multicolumn{1}{c|}{}&
\multicolumn{1}{c}{}&
\multicolumn{1}{c|}{(MeV)}&
\multicolumn{1}{c}{Expt.}&
\multicolumn{1}{c}{HFB-u}&
\multicolumn{1}{c}{HFB}&
\multicolumn{1}{c}{SD}&
\multicolumn{1}{c}{SDPF}\\
\hline
$^{24}$Na & $\frac{3}{2}^{+}$           &  6.96  & $100\pm20$ &54.0&122.3&149.7&91.1
\\
$^{25}$Mg & $0^{+}$ & 7.33 & $480\pm70$    &99.6&379.5&696.4&490.7
\\
$^{26}$Mg  & $\frac{5}{2}^{+}$           & 11.09 &$50\pm10$ & 8.7 & 46.1 & 78.9  & 54.2
\\
$^{27}$Mg & $0^{+}$           & 6.45  &$210\pm80$ &136.4&319.8&578.4 &226.0\\
\hline
\end{tabular}}
\end{table}

In Table \ref{tab:resonance}, we list the experimental and calculated values of $D_0$ obtained by using SDM and mean-field models. The $D_0$ values corresponding to $sd$ model space \cite{Karampagia2018} are overestimated by $50\%$ for $^{24}$Na, $^{25,26}$Mg. For the case of $^{27}$Mg, the $D_0$ value corresponding to $sd$ model space is off by more than $100\%$ whereas their NLDs are found to be in close agreement with the experimental low-lying levels (Fig. \ref{nld}). This is in contrast to the results corresponding to $sd$-$pf$ model space which explain the $D_0$ value quite well within the experimental error. This may be due to incomplete low-energy levels known so far for $^{27}$Mg. Future measurements for the level structure of $^{27}$Mg up to neutron separation energy would be crucial to resolve this issue. The HFB-u calculations underestimate the measured $D_0$ values significantly. We have also examined the influence of allowed cross-shell excitations in our calculations. The $D_0$ values obtained with $E_{gs}$ corresponding to (0,2)$\hbar\omega$ excitations to $pf$-shell are underestimated by $\sim 10 \%$ in $^{24}$Na and by $\sim 20-30 \%$ in Mg isotopes. This suggests that the further inclusion of excitations to $pf$-shell may not alter the results drastically. The predicted values of $D_0$ in $^{25}$Na and $^{26}$Na
are 46.8 keV and 45.6 keV, respectively, using $sd$-$pf$ model space which are much smaller as compared to their respective values 85.1 keV and 118.9 keV using $sd$ model space \cite{Karampagia2018}.

\begin{figure}[!htb]
\begin{center}
\includegraphics[width=0.79\textwidth]{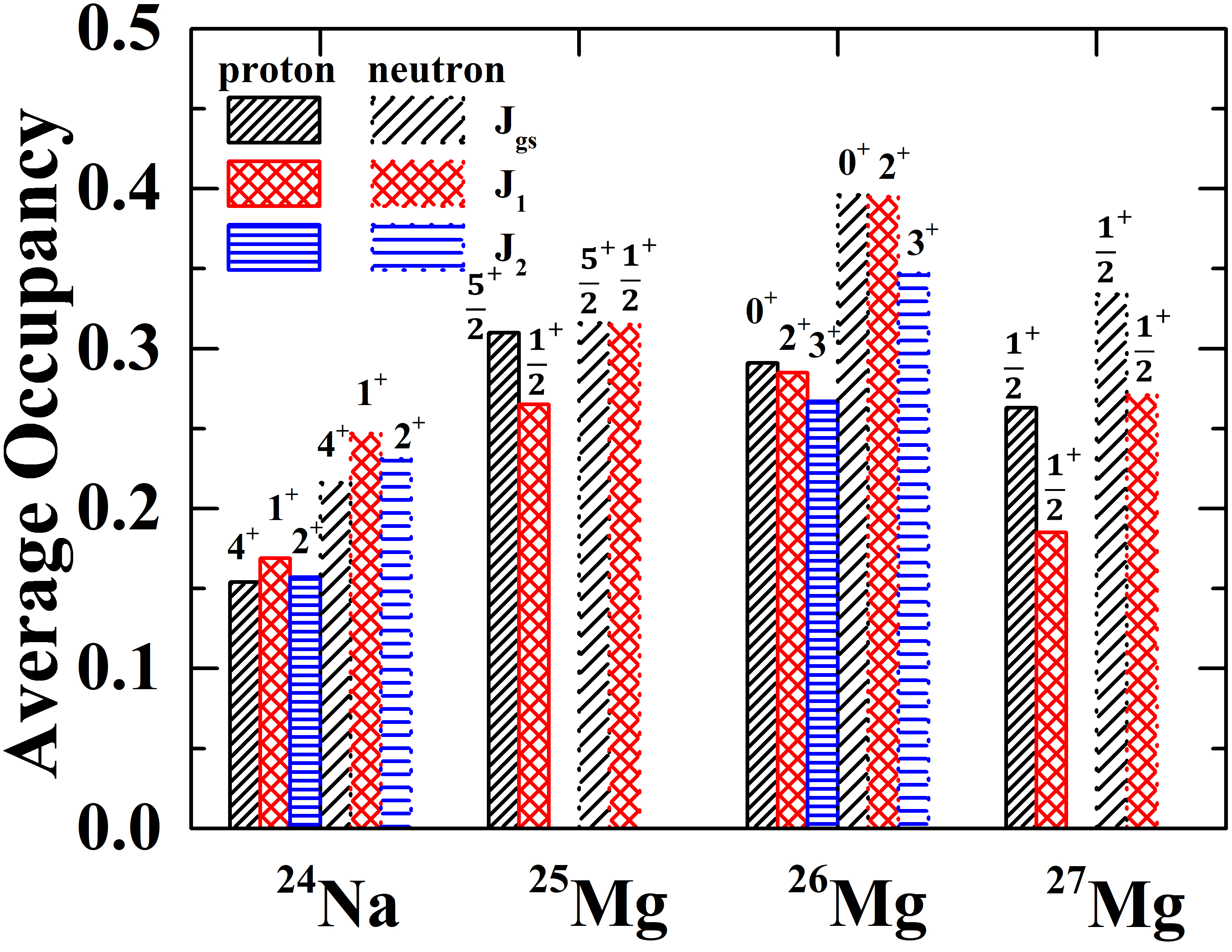}
\end{center}
\vspace{-0.5cm}
\caption{Average occupancies of protons and neutrons in $pf$-shell for the ground states (J$_{gs}$) as well as the excited states (J$_{1}$, J$_{2}$) relevant for the calculations of $D_0$. The shell model calculations have been performed using SDPF-MU interaction corresponding to (0,4) $\hbar\omega$ excitations in  $pf$-shell.}\label{occu}
\end{figure}

We present in Fig. \ref{occu}, the ground-state average occupancy of $pf$-shell for the protons and neutrons in the nuclei considered. Though, the sum of the proton and neutron occupancies of $pf$-shell come close to only $0.5$, this configuration mixing lowers the ground state energies by $5-8$ MeV. The resulting level densities increase in such a way that the values of $D_0$ get significantly modified and become closer to the experimental data. A small fraction of $pf$-shell's average occupancy is proven to be instrumental in explaining the experimental data. To examine further, we calculate the average $pf$-shell occupancy of excited states with spins and parities relevant to $D_0$. All these occupancies are obtained with (0,4) $\hbar\omega$ excitations in $pf$-shell, except in the case of excited ${1/2}^+$ state in $^{27}$Mg for which the calculations could be performed only with (0,2) $\hbar\omega$ excitations in $pf$-shell.

We list in Table \ref{tab:occup}, the average occupancies with (0,4) $\hbar\omega$ excitations in $pf$-shell for each orbit of protons and neutrons in $sd$-$pf$ model space for $^{24}$Na and $^{25}$Mg. The superscript $f$ denotes the full space calculations wherever possible. The average occupancies with (0,4) $\hbar\omega$ excitations in $pf$-shell come out to be very close to the full space calculations. On comparing these occupancies with those obtained with only $sd$ model space, the cross-shell excitations are predominately taking place from $d$-orbits of $sd$-shell to $f$-orbits in $pf$-shell. For example, the average occupancy of protons and neutrons in $d$-orbits for $^{24}$Na changes from 2.683 to 2.469, and 4.628 to 4.366, respectively, on going from $sd$ to $sd$-$pf$ model space. This leads to occupancy of protons and neutrons in $f$-orbits as 0.130 and 0.182, respectively. This may be plausibly due to the more attractive nature of matrix elements between $d_{5/2,3/2}$ and $f_{7/2,5/2}$ orbits in comparison to the coupling between an $s_{1/2}$ orbit and the $p_{3/2,1/2}$ orbits \cite{Casten2001, Otsuka2022}. \par

\begin{sidewaystable}
\centering
\caption{\label{tab:occup} Shell model average occupancies of the orbitals in $sd$-$pf$ model space for $^{24}$Na and $^{25}$Mg. The shell model Hamiltonian has been diagonalized with SDPF-MU interaction using KSHELL. The results are listed for the ground states as well as the excited states used in $D_0$ calculations within (0,4) $\hbar\omega$ excitations. $^{24}$Na$^f$ and $^{25}$Mg$^f$ present average occupancies obtained using the full space calculations (wherever possible).}
\resizebox{\textwidth}{!}{%
\begin{tabular}{c c c c c c c c c c c c c c c c }
\hline
\multicolumn{1}{c}{Nucleus}&
\multicolumn{1}{c}{J$^\pi$}&
\multicolumn{13}{c}{Shell model occupancies}\\
\cline{3-16}
&& $\pi 1d_{5/2}$ & $\pi 1d_{3/2}$&$\pi 2s_{1/2}$& $\pi 1f_{7/2}$ & $\pi 1f_{5/2}$ & $\pi 2p_{3/2}$ & $\pi 2p_{1/2}$ & $\nu 1d_{5/2}$ & $\nu 1d_{3/2}$&$\nu 2s_{1/2}$& $\nu 1f_{7/2}$ & $\nu 1f_{5/2}$ & $\nu 2p_{3/2}$ & $\nu 2p_{1/2}$\\

\hline
$^{24}$Na & $4^+$ & 2.065 & 0.404 & 0.377 & 0.077 & 0.053 & 0.015 & 0.009 & 3.802 & 0.564 & 0.418 & 0.121 & 0.061 & 0.023 & 0.011 \\
& $1^+$ &1.813 & 0.477 & 0.540 & 0.082 & 0.054 & 0.021 & 0.012 & 3.201 & 0.908 & 0.643 & 0.133 & 0.065 & 0.033 & 0.016 \\
& $2^+$ & 1.862 & 0.477 & 0.506 & 0.073 & 0.053 & 0.019 & 0.012 & 3.176 & 0.867 & 0.727 & 0.121 & 0.062 & 0.033 & 0.015\\
$^{24}$Na$^f$ & $4^+$ & 2.067 & 0.401 & 0.378 & 0.077 & 0.054 & 0.015 & 0.009  & 3.805 & 0.561 & 0.416 & 0.121 & 0.061 & 0.023 & 0.011 \\
& $1^+$ &1.799 & 0.521 & 0.501 & 0.084 & 0.061 & 0.021 & 0.013 & 2.977 & 1.067 & 0.687 & 0.143 & 0.068 & 0.042 & 0.016 \\
& $2^+$ &1.757 & 0.569 & 0.503 & 0.082 & 0.051 & 0.024 & 0.013 & 3.154 & 0.732 & 0.872 & 0.128 & 0.061 & 0.036 & 0.017\\
$^{25}$Mg & ${5/2}^+$ & 2.548 & 0.735 & 0.408 & 0.169 & 0.093 & 0.032 & 0.016 & 3.515 & 0.704 & 0.464 & 0.174 & 0.089 & 0.036 & 0.017\\
& ${1/2}^+$ & 2.390 & 0.769 & 0.576 & 0.127 & 0.082 & 0.037 & 0.019 & 3.132 & 0.890 & 0.663 & 0.164 & 0.089 & 0.041 & 0.021\\
$^{25}$Mg$^f$ & ${5/2}^+$ & 2.534 & 0.731 & 0.422 & 0.171 & 0.094 & 0.032 & 0.016 & 3.485 & 0.719 & 0.472 & 0.179 & 0.091 & 0.037 & 0.017\\
\\
\hline
\end{tabular}}
\end{sidewaystable}

\begin{table}[!htb]
\caption{\label{tab:ctm} The parameters (in MeV) $E_0$ and T of the constant temperature model, derived from fitting the constant temperature formula to the level density calculated using the moments method, for all available $J$ values, at the energy interval $E_x=0-10$ MeV, for all the considered nuclei.}
\centering
\resizebox{0.66\textwidth}{!}{%
\begin{tabular}{c c c c c c c c c}
\hline
Nucleus & \multicolumn{2}{c}{$sd$ \cite{Karampagia2018}} & \multicolumn{2}{c}{$sd$-$pf$} & \multicolumn{2}{c}{HFB-u} & \multicolumn{2}{c}{HFB} \\
\cline{2-9}
& E$_0$ & T & E$_0$ & T & E$_0$ & T & E$_0$ & T \\
\cline{2-5}
\hline
$^{24}$Na & $-8.54$ & 3.58  & $-11.26$ & 3.40 & $-2.55$ & 2.00 &$-2.67$& 2.35 \\
$^{25}$Mg & $-4.75$ & 3.15 & $-6.66$ & 3.11 & $-4.94$ & 2.53 & $-3.52$ & 3.07\\
$^{26}$Mg & $-3.74$ & 3.11 & $-7.14$  & 3.14 & $-1.17$ & 1.88 & 0.2 & 2.19\\
$^{27}$Mg & $-7.26$ & 3.41 & $-14.11$ & 3.38  & $-3.18$ & 2.03 & 1.40 & 1.55\\
\hline
\end{tabular}}
\end{table}

\begin{figure}[!htb]
\vspace{-0.3cm}
\begin{center}
\includegraphics[width=0.7\textwidth]{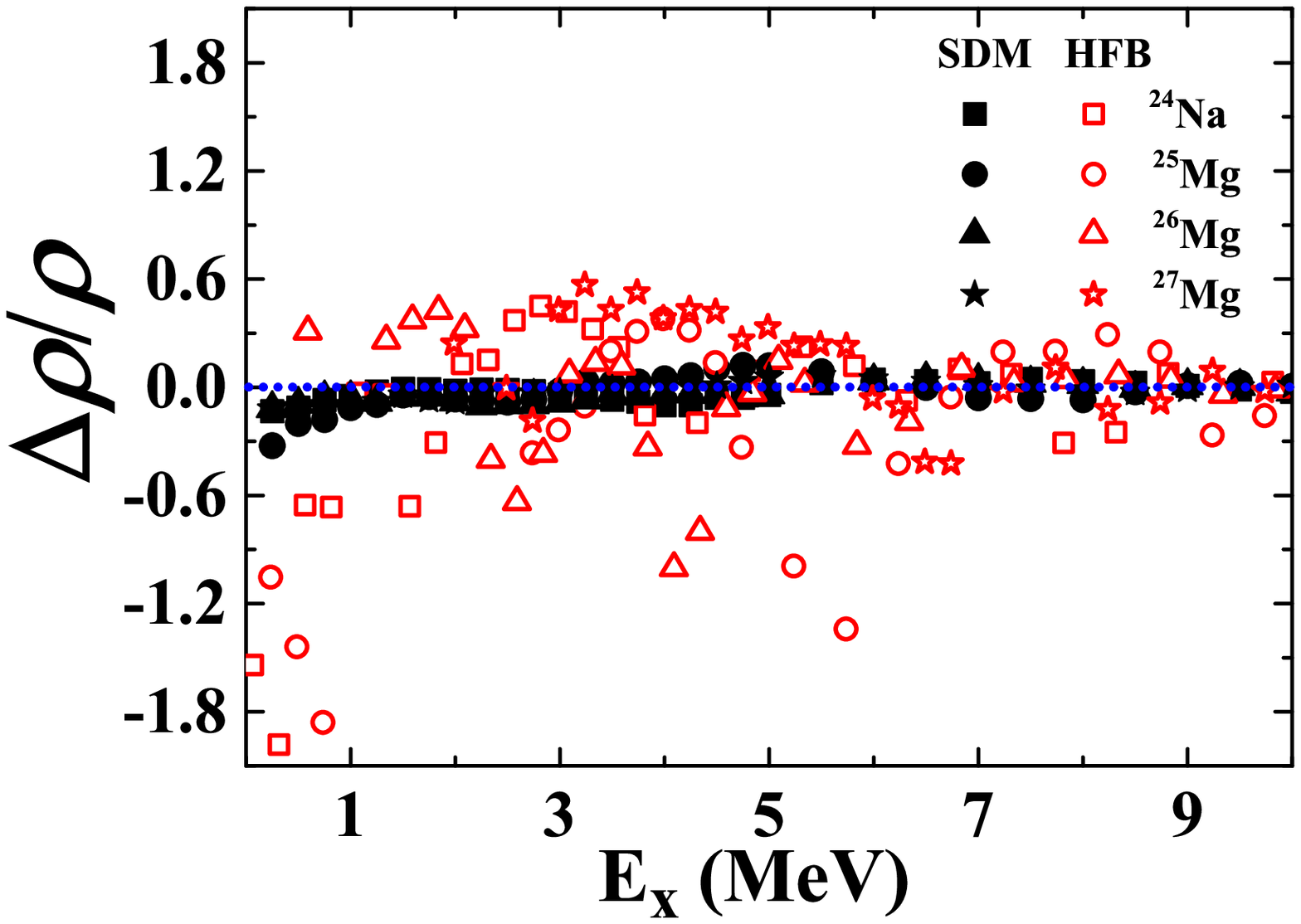}
\end{center}
\vspace{-0.5cm}
\caption{Relative deviations $\Delta \rho /\rho = (\rho-\rho_{CTM})/\rho$ as a function of excitation energies ($E_x$), with $\rho$ corresponding to SDM (solid symbols) and HFB (hollow symbols), and $\rho_{CTM}$ being the fitted NLDs from constant temperature model.}\label{shift}
\end{figure}

The average occupancies of $pf$-shell for the excited states near the neutron separation energies ($6-12$ MeV) remain nearly equal to those for the corresponding ground-state, indicating the validity of constant temperature model up to $S_n$ \cite{Gilbert1965} (see Fig. \ref{occu}).
We therefore fit the level densities, shown in Fig. \ref{nld}, to the constant temperature model given as:
\begin{equation}
\rho_{CTM}({E_x}) = \frac{1}{T}exp\Bigg( \frac{{E_x}-E_{0}}{T}\Bigg)  
\end{equation}
We list the values of $E_0$ and $T$ in Table \ref{tab:ctm} obtained by fitting the NLDs from different models to the constant temperature model over the excitation energies up to 10 MeV. The correlation between the parameters $E_0$ and $T$ is almost unity. The values of $\chi^2$ per degrees of freedom for the HFB are found to be higher than those for the SDM. We display the relative deviation of level densities from constant temperature model, $\Delta \rho /\rho = (\rho-\rho_{CTM})/\rho$ as a function of excitation energies ($E_x$) in Fig. \ref{shift}, for SDM in $sd$-$pf$ model space (solid symbols) and HFB (hollow symbols) results.
The SDM results are very close to $\Delta \rho /\rho =0$, whereas, the HFB results show relatively larger deviations. It is further found that the level densities obtained from SDM start to deviate from constant temperature model for the excitation energies larger than 10 MeV \cite{Gilbert1965}. The SDM results for $sd$ model space also follow the constant temperature model as can be seen from the Ref. \cite{Karampagia2018}. 


\section{Conclusion}
We calculate spin and parity dependent nuclear level densities (NLDs) using spectral distribution method (SDM). This method is numerically quite efficient and also involves the contributions from configuration mixing through a residual interaction. The SDM calculations are performed for $^{24}$Na and $^{25,26,27}$Mg nuclei using large scale $sd$-$pf$ model space with SDPF-MU interaction which enables one to incorporate the $sd$ to $pf$ cross-shell excitations.
The required ground state energies are obtained with (0,4)$\hbar\omega$ excitations to $pf$-shell which are found to be very close to the possible full space calculations. Our SDM NLDs are employed to calculate the s-wave neutron resonance spacing ($D_0$), which is important in estimating the astrophysical reaction rates. 

The calculated values of $D_0$ using $sd$-$pf$ model space are found to be in much better agreement with the experimental data in comparison to those obtained with only $sd$ model space. This is due to the presence of cross-shell excitations; though the average occupancy of $pf$-shell does not reach beyond 5$\%$ of the valence nucleons for the nuclei considered. Such small average occupancies of $pf$-shell strikingly reduces the $D_0$ significantly. The $D_0$ values for $^{25,26}$Na have also been predicted using $sd$-$pf$ model space. It will also be interesting to consider the influence of $1p$ orbitals to $sd$-shell excitations on the level densities and neutron resonance spacing. 
In the present calculations, we have considered more than one major shell, which may introduce spurious states due to center-of-mass motion \cite{Gloeckner1974}. There are systematic methods of getting rid of these unphysical states, in particular in application to the level density \cite{Horoi2007}. In the current study, we have not considered this correction; nevertheless, owing to the low $pf$-shell occupancies, we anticipate it to be notably smaller.



\section*{Acknowledgments}
TG acknowledges Council of Scientific and Industrial Research (CSIR), Government of India for fellowship Grant No. 09/489(0113)/2019-EMR-I. BM acknowledges the financial support from the Croatian Science Foundation and the \'Ecole Polytechnique F\'ed\'erale de Lausanne, under the project TTP-2018-07-3554 ``Exotic Nuclear Structure and Dynamics", with funds of the Croatian-Swiss Research Programme. GS and BKA acknowledge partial support from the SERB, Department of Science and Technology, Government of India with grant no. SIR/2022/000566 and CRG/2021/000101, respectively.
\section*{REFERENCES}
\bibliography{mybibfile}
\end{document}